\documentclass[a4paper]{jpconf}
\usepackage{graphicx}
\usepackage{placeins}
\begin{document}
\title{Disk storage management for LHCb based on Data Popularity estimator}

\author{Mikhail Hushchyn}

\address{Yandex School of Data Analysis, Moscow, Russian Federation\newline 
Moscow Institute of Physics and Technology, Moscow, Russian Federation}

\ead{mikhail91@yandex-team.ru}

\author{ Philippe Charpentier}

\address{CERN - European Organization for Nuclear Research}

\ead{Philippe.Charpentier@cern.ch}

\author{Andrey Ustyuzhanin}

\address{Yandex School of Data Analysis, Moscow, Russian Federation\newline
National Research University Higher School of Economics (HSE), Moscow, Russian Federation\newline
NRC "Kurchatov Institute", Moscow, Russian Federation\newline
Moscow Institute of Physics and Technology, Moscow, Russian Federation}

\ead{anaderi@yandex-team.ru}

\begin{abstract}
This paper presents an algorithm providing recommendations for optimizing the LHCb
data storage. The LHCb data storage system is a hybrid system. All datasets are
kept as archives on magnetic tapes. The most popular datasets are kept on
disks. The algorithm takes the dataset usage history and metadata
(size, type, configuration etc.) to generate a recommendation report. This
article presents how we use machine learning algorithms to predict future data
popularity. Using these predictions it is possible to estimate which datasets
should be removed from disk. We use regression algorithms and time series
analysis to find the optimal number of replicas for datasets that are kept on
disk. Based on the data popularity and the number of replicas optimization, the
algorithm minimizes a loss function to find the optimal data distribution. The loss function represents all requirements for data distribution in the data storage system. We demonstrate how our algorithm helps to save disk space and to reduce waiting times for jobs using this data.
\end{abstract}

\section{Introduction}
The LHCb collaboration is one of the four major experiments at the Large Hadron
Collider at CERN. The detector, as well as the Monte Carlo simulations of
physics events, create vast amount of data every year. This data is kept on disk
and tape storage systems. Disks are used  for storing data used by physicists
for analysis. They are much faster than tapes, but are way more expensive and
hence disk space is limited. Therefore it is highly important to identify which
datasets should be kept on disk and which ones should only  be kept as archives
on tape. Currently, the data volumes on disk and tape are about 10.5 PB and 1.5 PB respectively. The algorithm presented here is designed to select the datasets which may
be used in the future and thus should remain on disk. Input information to the
algorithm are the dataset usage history and dataset metadata (size, type, configuration etc.).

The algorithm consists of three separate modules. The first one is the
Data Popularity Estimator. This module predicts the dataset future popularity by applying a machine learning algorithm to the algorithm's input information. The data popularity represents the probability for a dataset to be useful in future. Based on data popularity it is possible to identify which datasets can be removed from disk. 

The second module is the Data Intensity Predictor. This module is needed to
predict the future usage intensity of each dataset. Time series analysis and regression algorithms are used to make these predictions. Input information for this module is the dataset usage history. 

The third module is the Data Placement Optimizer. In this module the data
popularity and the predicted future usage intensities are used to estimate which
datasets should be kept on disk and how many replicas they should have. For this
purpose a loss function minimization problem is solved. The loss function
represents all requirements for data distribution in the data storage
system.

These three modules are described in detail in the following sections. In the
results section we then show a comparison of our algorithm with a simple Last Recently Used (LRU) algorithm.

\section{Related works}
A Data Management Algorithm for hybrid hard disk drive (HDD) + solid-state drive (SSD) data storage system is described in [2]. The authors presents a method that shuffles datasets across storage tiers to optimize the data access performance. The method uses Markov chains[1] to predict the popularity of dataset accesses. The dataset placement optimization problem is solved based on the dataset accesses popularity. 

A Popularity-Based Prediction and Data Redistribution Tool for the ATLAS
Distributed Data Management is presented in [3,4]. The authors use artificial
neural networks (ann)[1] to predict possible dataset accesses in the near-term
future based on the dataset usage history. Then these predictions are used to redistribute data on the grid, i.e., adding and removing replicas. 

A feature of our study is that dataset usage history in LHCb has a rather low
statistics. The Data Management Algorithm from [2] needs more statistics for a good
performance. The artificial neural networks from articles [3,4] are too
complicated for our data and as a consequence an overfitting problem[1] may occur. 

\section{Input information}
Dataset usage history and metadata are used as input information to the
algorithm. In this study we use weekly dataset usage counters collected over the
last two years. Dataset usage history represents as time series of 104
points. Each point represents the number of dataset usages during one week
(i.e. the number of files accessed by Grid jobs divided by the number of files
in the dataset).

The dataset metadata contains additional dataset information likes: the origin, the detector configuration,  the file type, the
data type (Monte Carlo simulations or real data), the event type, the creation
week, the first usage week, the last usage week, the size for one replica, the
total size of occupied disk space, the number of replicas on disk and some others. 

The algorithm takes as input a file which contains the dataset usage history and the dataset metadata. This file comes from the file catalogue.

\section{Data Popularity Estimator}
The Data Popularity Estimator module uses a classifier to calculate the data
popularity. The classifier is a supervised machine learning algorithm[1] and
consists of several steps. The following subsections describe each step of data popularity estimation. 

\subsection{Dataset labels}
As the classifier is a supervised machine learning algorithm, each dataset
should be labelled as popular or unpopular. The time series of dataset usage history are very sparse, therefore
the last 26 weeks of usage history are used to label the data. If a dataset has
not be used during the last 26 weeks we label it as unpopular and assign it a
label value "1". Otherwise, the dataset is labelled as popular with a label
value "0". This label defines the class of the dataset (0 for popular, 1 for
unpopular). The figures 1 and 2 show each the time series of one dataset of each class.

\begin{figure}[h]
\begin{center}
\begin{minipage}{16pc}
\includegraphics[width=16pc]{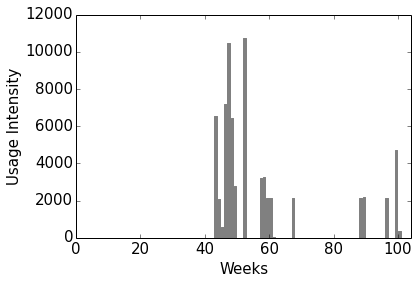}
\caption{\label{label}Time series with label "0".}
\end{minipage}\hspace{2pc}%
\begin{minipage}{16pc}
\includegraphics[width=16pc]{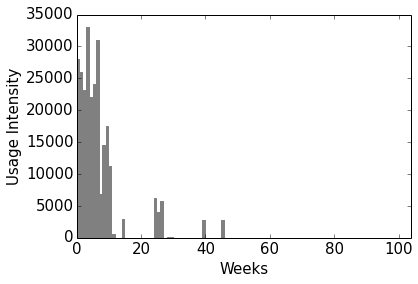}
\caption{\label{label}Time series with label "1".}
\end{minipage} 
\end{center}
\end{figure}

\subsection{Data preprocessing}
The dataset metadata are used as input parameters for a classifier. Some new
parameters are computed and used in the analysis as well as the existing
ones. These factors describe the shape of the time series of the dataset usage
history. While the last 26 weeks of the time series are used to label the
datasets, the first 78 weeks are used to compute these new parameters. These parameters are \textit{nb\_peaks}, \textit{last\_zeros}, \textit{inter\_max}, \textit{inter\_mean}, \textit{inter\_std}, \textit{inter\_rel}, \textit{mass\_center}, \textit{mass\_center\_sqrt}, \textit{mass\_moment} and \textit{r\_moment}.

\textit{Nb\_peaks} is the number of weeks during which a dataset has been
used. \textit{Last\_zeros} is the number of weeks since when the dataset was
last used. \textit{Inter\_max}, \textit{inter\_mean}, \textit{inter\_std} are
the maximum value, the mean value and the standard
deviation of the number of weeks between consecutive weeks of
usage. \textit{Inter\_rel} is the ratio of the \textit{inter\_std} and
\textit{inter\_mean} values. \textit{Mass\_center} is the center of gravity of a
time series for a dataset, where the "mass" is the number of accesses to the
dataset for each week. \textit{Mass\_center\_sqrt}, \textit{mass\_moment} and \textit{r\_moment} are similar to \textit{mass\_center}, but "mass" and "coordinate" have different degrees.

These parameters significantly increase the classifier's quality.

\subsection{The classifier training}
The new parameters, the dataset metadata and their labels are used to train a
\textit{Gradient Boosting Classifier}[1]. All datasets are split into two equal
halves, i.e. half the datasets goes to the first halve, the other to the second one. The classifier is trained on one half of the datasets and then is used
to predict probabilities to have label "1" for the second half of the
datasets. The figure 3 shows the distribution of the probabilities for each class of datasets.

\subsection{Popularity estimation}
The probability described previously is then transformed into a popularity
estimator such that the popularity for datasets which have label "1" is
uniform. The closer the popularity is to 1 the higher is the probability that it
will be unused in the future. In this sense it is rather an 'unpopularity'
estimator. The figure 4 represents the distribution of the popularity for each dataset class.

\begin{figure}[h]
\begin{center}
\begin{minipage}{14pc}
\includegraphics[width=14pc]{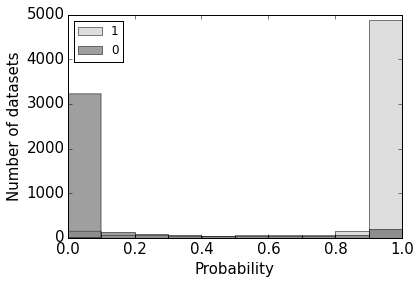}
\caption{\label{label}Distributions of the probability to have label "1" for each dataset class.}
\end{minipage}\hspace{2pc}%
\begin{minipage}{14pc}
\includegraphics[width=14pc]{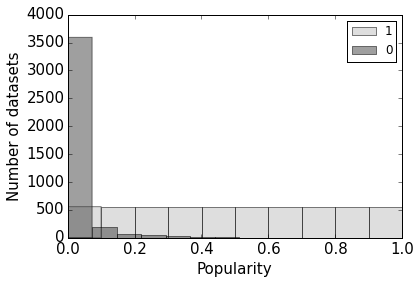}
\caption{\label{label}Distributions of the popularity for each dataset class.\newline}
\end{minipage} 
\end{center}
\end{figure}

\section{Data Intensity Predictor} 
The data popularity represents the probability that a dataset will be unused in the future. Another important
feature is predicted dataset usage intensity. There is a number of time series analysis
algorithms that predict future values of time series. Since time series in this study have lack of statistics, parametric
models such as polynomial regression, autoregression, ARMA and ARIMA models,
artificial neuron networks (ANNs) and others are not suitable. This section
shows how to use two non-parametric models to predict the dataset usage intensities. These models are Nadaraya-Watson kernel smoothing[1] and rolling mean values[1].

\subsection{Nadaraya-Watson kernel smoothing}

Let points $(x_{1}, y_{1}), (x_{2}, y_{2}), ..., (x_{l}, y_{l})$ represent a
time series and $X^{l}=\{x_{1}, x_{2}, ..., x_{l}\}$. Then, the
\textit{Nadaraya-Watson} equation for kernel smoothing is:
\begin{equation}
\hat{y}_{h}(x; X^{l})=\frac{\sum_{i=1}^{l}y_{i}K(\frac{\rho(x,x_{i})}{h})}{\sum_{i=1}^{l}K(\frac{\rho(x,x_{i})}{h})},
\end{equation}
where 

$\hat{y}_{h}(x; X^{l})$ is the time series value at $x$ after kernel smoothing of $X^{l}$ values,

$K(\frac{\rho(x,x_{i})}{h})=exp(-\frac{(x-x_{i})^{2}}{2h^{2}})$ is the RBF smoothing kernel,

$h$ is the smoothing window width.
\newline
\newline
For the smoothing window width optimization the \textit{Leave-One-Out}[1] method was applied:
\begin{equation}
LOO(h, X^{l})=\sum_{i=1}^{l}(\hat{y}_{h}(x_{i}; X^{l}\setminus{\{x_{i}\}})-y_{i})^{2} \mapsto \min_{h}
\end{equation}
The \textit{Nadaraya-Watson} equation for kernel smoothing with \textit{LOO}
smoothing window width optimization is applied to time series of dataset usage
history. The maximum smoothing window width is 30 weeks. The figure 5 shows an
example of time series after this smoothing is applied.

\subsection{Rolling mean values calculation}
On the next step rolling mean values are calculated for additional smoothing of
time series of dataset usage history. Let points $(x_{1}, y_{1}), (x_{2},
y_{2}), ..., (x_{l}, y_{l})$ represent a time series after the kernel
smoothing. Then, rolling mean values are defined as:
\begin{equation}
\hat{y}_{k}=\frac{\sum_{i=k-w}^{k}y_{i}}{w}
\end{equation}
where $w$ is the width of the moving window.

The window width is chosen such that 90\% of all time series with equal \textit{nb\_peaks} values have \textit{inter\_max} values less than the window width.

The rolling mean value at moment $x_{i}$ represents the dataset usage intensity
at that moment. The simplest way to predict future dataset usage intensity is to
take dataset usage intensity on last observation as future one. An example of calculated rolling mean values and predicted dataset usage intensity are shown on the figure 5.

\begin{figure}[h]
\begin{center}
\begin{minipage}{14pc}
\includegraphics[width=14pc]{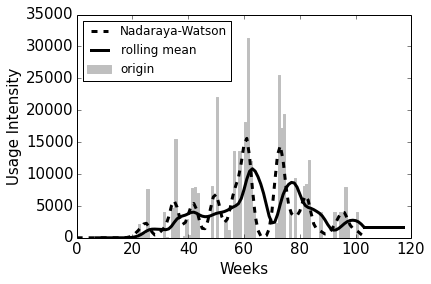}
\caption{\label{label}Example of time series after \textit{Nadaraya-Watson} kernel smoothing and rolling mean values calculation.}
\end{minipage}\hspace{2pc}%
\begin{minipage}{14pc}
\includegraphics[width=14pc]{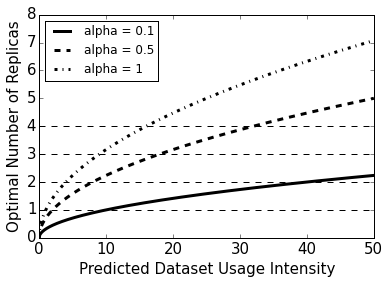}
\caption{\label{label}Dependence of optimal number of dataset replicas ($Rp$) from its predicted usage intensity ($I$) and $\alpha$.}
\end{minipage}
\end{center}
\end{figure}

\section{Data Placement Optimizer}
This section describes how one can estimate which dataset should be kept on disk
and how many replicas they should have using the popularity and the predicted
usage intensity for this dataset. Since disk space is more expensive than tapes,
we would like to take a minimum of disk space. But on the other hand it is
highly undesirable to remove from disk datasets which will be used in
future. Additionally, we would like to create more replicas for the most popular
datasets in order to reduce their average access time.

The requirements above are represented by the following loss function:
\begin{equation}
L=C_{disk}\sum_{i}^{n}S_{i}(Rp_{i}+\alpha\frac{I_{i}}{Rp_{i}})\delta_{i}+
C_{tape}\sum_{i}^{n}S_{i}(1-\delta_{i})+
C_{miss}\sum_{i}^{n}S_{i}m_{i},
\end{equation} 

$C_{disk}$ - cost of 1 Gb disk storage,

$C_{tape}$ - cost of 1 Gb tape storage,

$C_{miss}$ - cost of restoring 1 Gb data from tape to disk,

$\alpha$ - penalty for low number of replicas,

$S_{i}$ - size of one replica of $i^{th}$ dataset,

$Rp_{i}$ - number of replicas of $i^{th}$ dataset,

$I_{i}$ - predicted usage intensity of $i^{th}$ dataset;

$\delta_{i}$ is equal to 1 if $i^{th}$ dataset is on disk, otherwise it is 0;

$m_{i}$ is equal to 1 if $i^{th}$ dataset was restored from tape to disk.
\newline

The first term of the loss function represents the cost of storage of the datasets on disk. The second term is the cost of storage of the datasets on tape. The last term is the cost of mistakes, when a dataset was removed from disk but then is used. 

The expression in brackets in the first term of the loss function is used to find the optimal number of replicas for datasets on disk based on predicted usage intensities. The optimal number of replicas for a dataset with predicted usage intensity of $I_{i}$ and for the value $\alpha$ is
\begin{equation}
Rp_{i\_optimal}=\sqrt[]{\alpha I_{i}},
\end{equation}
The figure 6 shows how the optimal number of replicas for a dataset depends on
its predicted usage intensity and alpha value. For example, suppose the predicted usage intensity for a dataset is $I = 10$ usages per week and $\alpha = 0.5$. Then $Rp_{optimal}=\sqrt{\alpha I}=\sqrt{0.5*10}=2.24\approx2$ replicas.

The $\delta_{i}$ value in the loss function depends on the data popularity threshold value. Datasets with popularities equal to or higher than this threshold value are removed from disk ($\delta_{i}=0$). The $m_{i}$ value is the product of the $1-\delta_{i}$ and the label of $i^{th}$ dataset (0 or 1). 

The loss function optimization consists in finding the data popularity threshold value and dataset optimum number of replicas that provide the minimum value of the loss function.

\section{Results}
\subsection{LRU algorithm}
In this article we compare our algorithm with the Last Recently Used
(LRU) algorithm. The LRU algorithm takes the last observations of the dataset
usage history and decides which dataset should be removed from disk. In this
study the first 78 weeks usage history time series are used as the algorithm
inputs. The last 26 weeks are used to measure the quality of the algorithm. Thus
if a data set was not used during the last $N$ weeks (from $78-N^{th}$ to
$78^{th}$ weeks), this dataset is removed from disk. The number of disk replicas
are not changed compared to the original number of replicas.

\subsection{Downloading time}
The following function is used to estimate the time of downloading of all
datasets by all users (the generic term 'downloading' is used to represent an
access to the dataset from a job):
\begin{equation}
T=\sum_{i=1}^{n}I_{i}^{*}S_{i}t_{disk}\alpha(Rp_{i})\delta_{i}+ 
\sum_{i=1}^{n}(K_{tape}+S_{i}t_{tape})m_{i}+
\sum_{i=1}^{n}I_{i}^{*}S_{i}t_{disk}m_{i}
\end{equation}

where $\alpha(Rp_{i})=0.05+\frac{1}{Rp_{i}}$

$t_{disk}$ - average time of downloading 1 Gb of data from disk,

$t_{tape}$ - average time of downloading 1 Gb of data from tape to disk,

$K_{tape}$ - constant time needed to restore a dataset from tape to disk,

$I_{i}^{*}$ - average number of downloading of a dataset per week,

$S_{i}$ - size of one replica of $i^{th}$ dataset,

$Rp_{i}$ - number of replicas of $i^{th}$ dataset,

$\delta_{i}$ is equal 1 if the $i^{th}$ dataset is on disk, otherwise it is 0,

$m_{i}$ (misclassification) is equal 1 if $i^{th}$ dataset has to be restored from tape to disk.
\newline

The first term of the downloading time equation represents the time of download
of all datasets from disk by all users. The second term represents the time
needed to restore from tape datasets
that were removed from disk due to an algorithm’s bad decision. The third term
represents the time of download of restored datasets by all users.The first 78
weeks of the dataset usage history time series are used as algorithms
inputs. The last 26 weeks are used to measure the quality of the algorithms and to estimate how many times the datasets were downloaded.

\subsection{Algorithms comparison}
Datasets which were created and first used earlier than $78^{th}$ week are used
to compare algorithms. The total number of datasets used for the comparison is 7375. In this paper we use rather pessimistic values of the parameters to emphasize that the disk space is highly limited. The following values of the parameters are used to optimize the loss function: $C_{disk}=100$, $C_{tape}=1$, $C_{miss}=2000$. The values of the parameters for the downloading time function are $t_{disk}=0.1$ hour/Gb, $t_{tape}=3$ hours/Gb and $K_{tape}=24$ hours. $C_{disk} \gg C_{tape}$ represents an idea that the disk space is limited. $C_{miss} \gg C_{disk}$ means the number of restored datasets should be minimal. $t_{tape} \gg t_{disk}$ and large $K_{tape}$ value show that a dataset restoring from tape to disk takes a lot of time.

Tables 1 and 2 show results for our algorithm with 4 maximum dataset
number of replicas and for the LRU algorithm. \textit{Downloading time ratio} is
the ratio of the downloading time after applying the algorithm to the original
downloading time. \textit{Saving space} column shows how much disk space can be
saved using this algorithm. \textit{Nb of wrong removings} column represents the
number of datasets which are proposed to be removed from disk but are then used
again in the future.

Both algorithms save about the same amount of disk space, but our algorithm has an extremely low number of mistakes. The tables show that our algorithm
with 4 maximum dataset number of replicas slightly decreases the download time.

Table 3 demonstrates that for a maximum number of replicas of 7 our algorithm helps
to save up to 40\% of disk space and decreases the downloading time by up to 30\%.

\FloatBarrier
\begin{table}[h]
\caption{\label{label}Results for LRU algorithm.}
\begin{center}
\begin{tabular}{llll}
\br
N&Downloading time ratio&Saving space, \%&Nb of wrong removings\\
\mr
1&1.33&63&1973\\
2&1.28&58&1659\\
5&1.4&50&1357\\
10&1.11&44&966\\
15&1.07&38&635\\
20&1.03&33&370\\
25&1.02&30&193\\
\br
\end{tabular}
\end{center}
\end{table}
\FloatBarrier

\FloatBarrier
\begin{table}[h]
\caption{\label{label}Results for our algorithm with 4 maximum numbers of replicas.}
\begin{center}
\begin{tabular}{llll}
\br
Alpha&Downloading time ratio&Saving space, \%&Nb of wrong removings\\
\mr
0&3.35&71&9\\
0.01&0.99&46&9\\
0.05&0.96&34&9\\
0.1&0.96&30&9\\
0.5&0.96&23&9\\
1&0.96&19&9\\
2&0.96&16&9\\
\br
\end{tabular}
\end{center}
\end{table}
\FloatBarrier

\FloatBarrier
\begin{table}[h]
\caption{\label{label}Results for our algorithm with 7 maximum numbers of replicas.}
\begin{center}
\begin{tabular}{llll}
\br
Alpha&Downloading time ratio&Saving space, \%&Nb of wrong removings\\
\mr
0&3.35&71&8\\
0.001&1.03&57&8\\
0.005&0.72&40&8\\
0.01&0.68&34&8\\
0.05&0.63&11&8\\
0.1&0.62&1&8\\
\br
\end{tabular}
\end{center}
\end{table}
\FloatBarrier

A python module implementing our algorithm and its web service can
be downloaded from [5]. Our study is performed by means of a Reproducible Experiment Platform[6] - environment for conducting data-driven research in a consistent and reproducible way.

\section{Conclusion}
In this paper, we presented a study of developing the algorithm for
disk storage management. The method presented here demonstrates how the algorithms of machine learning, regression and time series analysis can be used in data management of the LHCb data storage system. The results shows that our algorithm helps to save a significant amount
of disk space and reduce the average downloading time.

\end{document}